\documentclass[pra,reprint,amssymb,amsmath,aps,groupedaddress,superscriptaddress]{revtex4-1}
\usepackage[english]{babel}
\usepackage{graphicx,bm}
\usepackage{braket}

\usepackage{xcolor}

\newcommand{\Alpha}{\mathrm{A}}

\newcommand{\Eta}{\mathrm{H}}
\newcommand{\mathD}{\mathrm{D}}
\newcommand{\mathd}{\mathrm{d}}
\newcommand{\mathi}{\mathrm{i}}

\newcommand{\tmop}[1]{\ensuremath{\operatorname{#1}}}

\begin{document}

\title{Decoherence of nonrelativistic bosonic quantum fields}

\author{Marduk Bola{\~n}os}
\affiliation{Fakult{\"a}t f{\"u}r Physik, Universit{\"a}t
	Duisburg--Essen, Lotharstra{\ss}e 1-21, 47057 Duisburg, Germany}
\author{Benjamin A. Stickler}
\affiliation{Fakult{\"a}t f{\"u}r Physik, Universit{\"a}t
	Duisburg--Essen, Lotharstra{\ss}e 1-21, 47057 Duisburg, Germany}
\affiliation{QOLS, Blackett Laboratory, Imperial College London, London SW7 2BW, United Kingdom}
\author{Klaus Hornberger}
\affiliation{Fakult{\"a}t f{\"u}r Physik, Universit{\"a}t
	Duisburg--Essen, Lotharstra{\ss}e 1-21, 47057 Duisburg, Germany}

\begin{abstract} We present a generic Markovian master equation 
inducing the gradual decoherence of a bosonic quantum field. It leads
to the decay of quantum superpositions of field configurations,
while  leaving the Ehrenfest equations for both the field and the 
mode-variables invariant. We characterize the decoherence dynamics 
analytically and numerically, and show that the semiclassical field
dynamics is described by a linear Boltzmann equation in the functional
phase space of field configurations.
\end{abstract}

{\maketitle}

\section{Introduction}
Quantum systems with a large number of interacting constituents can
often be effectively described in terms of quantum fields. Examples
include degenerate quantum gases and fluids	 {\cite{Dalton,Schmiedmayer},
collective degrees of freedom in strongly correlated solid-state
systems \cite{Blasone,Greiner}, acoustic vibrations in superfluid helium {\cite{Harris,Schmitt,Marquardt}}, micromechanical
oscillators \cite{Riedinger,Ockeloen-Korppi}, and closely spaced chains
of harmonic oscillators realized, for example, in ion traps
{\cite{Monroe}} and superconducting circuits
{\cite{Superconducting}}. For such many-body systems a minimal and generic field-theoretic model that appropriately accounts for the decoherence dynamics toward a corresponding classical field theory is desirable.

The loss of quantum coherence and the emergence of classical
behavior in systems with a finite number of degrees of freedom have been extensively and successfully
studied using the framework of open quantum systems
{\cite{Breuer,GardinerI,*GardinerII,*GardinerIII}}. Understanding
these phenomena is of paramount importance for the development of
quantum technologies, since their performance is ultimately limited by
the coupling to the surrounding environment. Decoherence also plays a central role in
probing the physics at the quantum-classical border
\cite{Joos} as superpositions of increasingly macroscopic and complex
objects become experimentally accessible \cite{Arndt}.

For systems with a large number of interacting
constituents the combined decoherence dynamics become quickly intractable on an atomistic level. This calls for a field-theoretic description in terms
of collective modes. The open quantum dynamics of fields have
so far been formulated assuming a linear coupling with an environment
{\cite{Graham,Alicki,Calzetta,Albeverio,Anglin,Dalton}. While such schemes are adequate for
small fluctuations of the field amplitude, they cannot appropriately
describe the decoherence of macroscopic superpositions, since the obtained
rates become unrealistically large, growing above all bounds
\cite{Joos}.

In this Rapid Communication we introduce a generic Lindblad master equation for
nonrelativistic bosonic fields that describes their gradual
decoherence. That is, quantum superpositions of different field
configurations quickly decay into a mixture while classical
superpositions remain practically unaffected. We show that the semiclassical field
dynamics is described by a linear Boltzmann equation in the functional phase 
space of field configurations, which in the diffusion approximation 
reduces to a Fokker-Planck equation. The noise term in the master equation is minimal, in the sense that
it leaves the Ehrenfest equations for both the field and the mode
variables unaffected, while slowly increasing the field energy with a
state-independent rate.

\begin{figure}[t] \includegraphics[width=0.37\textwidth]{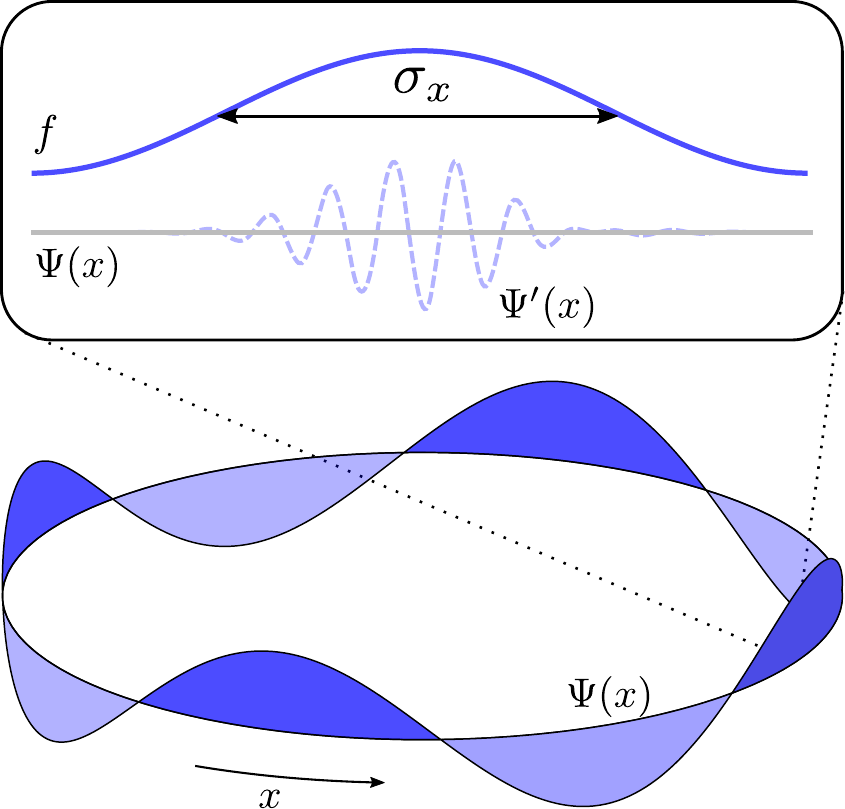}
	\caption{(Color online) The master equation (\ref{eq:master-equation}) decoheres the state of a bosonic quantum field $\Psi(x)$ by turning quantum superpositions into mixtures. The field dynamics can be viewed as 
	a random process in which the
	unitary evolution is interrupted by generalized measurements of the canonical field variables whose outcomes are discarded. These fictitious measurements have finite spatial resolution, as characterized by the spread function $f$ of width $\sigma_x$.\label{fig:measurement-diagram}}
\end{figure}

\section{{Field-theoretic master equation}}

We consider a bosonic
scalar field confined to a one-dimensional region of length $L$, subject to periodic
boundary conditions (the generalization to higher
dimensions is straightforward). In the Schr{\"o}dinger picture, its
quantum dynamics is described by the master equation
\begin{equation}
  \label{eq:master-equation} \dot{\rho}_t = - \frac{\mathi}{\hbar}
[\Eta, \rho_t] + \gamma \! \int_0^L \! \frac{\mathd x}{L} \! \int \!
\mathd^2 \xi \, g (\xi) \! \left[ \mathrm{U}_x (\xi) \rho_t \mathrm{U}_x^{\dagger} (\xi) - \rho_t \right].
\end{equation} The first term describes the unitary dynamics of the
field, as determined by the Hamiltonian $\Eta$, while the second term
gives rise to the decoherence of the field. The latter involves
unitary phase-space translation operators $\mathrm{U}_x(\xi)$ acting on
the field amplitude and its canonical conjugate momentum in the vicinity of position $x$. Combining
the field variables into the complex field $\Psi (x)$, so that $[\Psi (x),
\Psi^{\dagger} (y)] = \delta (x - y)$, these operators can be written as
\begin{equation}
\label{eq:displacement-op} \mathrm{U}_x (\xi) = \exp \left( \int_0^L
\mathd y\, f \left( \frac{y - x}{L} \right) \left[\xi \Psi^{\dagger} (y)
- \xi^{\ast} \Psi (y)\right] \right).
\end{equation} Here $f$ is a real, square-integrable, $L$-periodic spread function with maximum $f (0) = 1$ and width $\sigma_x/L$. The argument
$\xi$ of $\mathrm{U}_x$ is a complex random number whose associated probability
distribution $g (\xi)$ is assumed to be an even function of width $\sigma_g$.

From Eq.~{\eqref{eq:master-equation}} it follows that the purity $p_t = \tmop{Tr} (\rho_t^2)$ of 
any quantum state of the field decreases monotonically. { To see this we note that $\dot{p}_t = 2 \tmop{Tr} (\rho_t \dot{\rho}_t)$, and since $\tmop{Tr} (\rho_t [\Eta, \rho_t]) = 0$ and $\tmop{Tr} \left( \rho_t \mathrm{U}_x (\xi) \rho_t \mathrm{U}^{\dagger}_x
(\xi) \right) - p_t\ = - \frac{1}{2} \left\| \left[ \rho_t, \mathrm{U}_x (\xi)
\right] \right\|^2$, the purity decay rate is given by}
\begin{equation} 
\label{eq:monotonic-purity-decay}
\dot{p}_t = - \gamma \int_0^L \frac{\mathd x}{L} \int
\mathd^2 \xi \, g (\xi) \| [\rho_t, \mathrm{U}_x (\xi)] \|^2 < 0,
\end{equation} 
where $\| \Alpha \|^2 = \tmop{Tr}(\Alpha^{\dagger} \Alpha)$.
Therefore, the master equation induces the decay of any quantum
superposition of the field into a mixture. This loss of coherence will be
assessed quantitatively in Sec. IV.

The master equation can be interpreted as describing a compound Poisson process
with rate $\gamma$, in which the unitary evolution of the field is interrupted by 
generalized measurements of the canonical field variables whose outcomes are discarded
{\cite{Kossakowski,Cresser}}. Whenever a measurement occurs around a position 
$x \in [0, L]$ it affects an entire neighborhood of width $\sigma_x$, as illustrated 
in Fig.~\ref{fig:measurement-diagram}. The field degrees of freedom located at 
any point $y$ within this region experience a random phase-space kick of strength
$\xi f (y / L - x / L)$, in accordance with the Heisenberg principle. 
In the framework of generalized measurements one can thus construct a master equation inducing decoherence without specifying a physical mechanism for the incoherent dynamics, be it the interaction with a practically unobservable environment or, on a more fundamental level, a stochastic process augmenting the Schr\"odinger equation.

The operators in the master equation can be represented both in position space, in 
terms of the canonical field amplitude and its conjugate momentum, and in 
Fourier space, in terms of the mode variables. As will be shown the first representation is advantageous for semiclassical analysis, while the
second enables the analytic treatment of the decoherence dynamics. 

For definiteness, we take the bosonic field Hamiltonian as
\begin{equation}
\label{eq:Hamiltonian} \Eta = \frac{1}{2} \int_0^L \mathd x \left [
\frac{1}{\mu} \Pi^2 (x) + \mu \omega^2 \Phi^2 (x) + \mu v^2
[\partial_x \Phi (x)]^2 \right ].
\end{equation}
The field amplitude $\Phi$ and its canonically
conjugate momentum $\Pi$ are related to the complex field through 
\begin{equation}
\label{eq:complexField}
\Psi (x) = \sqrt{\frac{\mu\omega}{2\hbar}}\Phi (x) +  \frac{\mathi}{\sqrt{2\hbar\mu\omega}}\, \Pi (x).
\end{equation}
Here $\mu$ is the mass density of the field, $\omega$ is a
frequency, and $v$ is a speed.

For example, the Hamiltonian (\ref{eq:Hamiltonian}) can be used to model the collective dynamics of a chain of many trapped ions (see Ref.~\cite{Monroe}). For the continuum description to be valid, the width $\sigma_x/L$ of the spread function should be greater than the mean distance between the ions such that the discrete structure of the chain does not get resolved. Likewise, the width $\sigma_g$ of the kick distribution must be small enough such that the harmonic approximation remains valid.

The Hamiltonian (\ref{eq:Hamiltonian}) can be written in diagonal form as 
{$\Eta = \sum_{k\in K} \hbar \omega_k \mathrm{c}_k^{\dagger} \mathrm{c}_k$,}
with $\omega_k^2 = \omega^2 + v^2
k^2$ and $K=\{2\pi j/L\,|\, j\in\mathbb{Z}\}$, thus $k$ runs over all integer multiples of $2 \pi / L$. The 
bosonic mode operators
\begin{equation}
\label{eq:modeOperators}
\mathrm{c}_k = \sqrt{\frac{\mu\omega_k}{2\hbar}}\Phi_k +  \frac{\mathi}{\sqrt{2\hbar\mu\omega_k}}\, \Pi_k,
\end{equation}
satisfying the canonical commutation relations  $[\mathrm{c}_k, \mathrm{c}^\dagger_{k'}] = \delta_{k k'}$,
are defined in terms of the
(non-Hermitian) normal coordinates that diagonalize the Hamiltonian (\ref{eq:Hamiltonian}), $
\Phi_k = \int_0^L \mathd x e^{- \mathrm{i}  k x}
\Phi (x)/\sqrt{L}$ and $ \Pi_k = \int_0^L \mathd x e^{-
\mathrm{i}  k x} \Pi (x)/\sqrt{L}$.  In
the basis (\ref{eq:modeOperators}) the field operators are expressed as
\begin{equation}
  \label{eq:field-operator} \Psi (x) = \frac{1}{\sqrt{L}} \sum_{k\in K}
\left[ e^{\mathrm{i} k x} \, \Omega^+_k \mathrm{c}_k + e^{- \mathrm{i}
k x} \, \Omega^-_k \mathrm{c}_k^{\dagger}\right],
\end{equation}
with $\Omega_k^{\pm} = [(\omega/ \omega_k)^{1 / 2} \pm
(\omega_k/ \omega)^{1 / 2}]/2$.

{From a direct calculation using the Baker-Campbell-Hausdorff formula it follows} that the mode operators satisfy 
\begin{subequations}
\begin{eqnarray}
	[\Eta,\tmop{c}_k]&=&-\hbar\omega_k \tmop{c}_k,
	\\ \, 
	[\tmop{U}_x(\xi), \tmop{c}_k]&=& -f_k e^{-\mathi k x} (\xi \Omega_k^{+} - \xi^\ast \Omega_k^{-})\tmop{U}_x(\xi),
\end{eqnarray}
\end{subequations}
where $f_k = \int_0^L \mathd y \,
e^{\mathrm{i} k y} f \left({y}/{L} \right)/\sqrt{L}$ are the Fourier coefficients of $f$.
Since $g(\xi)$ is an even function the Ehrenfest equation for the mode operators $\partial_t\langle \tmop{c}_k \rangle_t = -\mathi \omega_k\langle\tmop{c}_k\rangle_t$  is unaffected by the incoherent part of the master equation (\ref{eq:master-equation}). The expectation values of the field amplitude $\langle\Phi(x)\rangle_t$ and its canonical momentum $\langle\Pi(x)\rangle_t$ therefore satisfy the field equations associated with the corresponding classical Hamiltonian.

The master equation {\eqref{eq:master-equation}} is most easily solved
in the basis of the Weyl operators $\mathD [\{ \eta_k \}] = \prod_{k\in K}\exp ( 
\eta_k \mathrm{c}_k^{\dagger} - \eta_k^{\ast} \, \mathrm{c}_k )$. For each mode variable with wave number $k$, they effect a phase-space displacement by the complex amplitude $\eta_k$.
In this representation, the state of the
field is encoded in the characteristic functional $\chi_t [\{\eta_k \}] = \tmop{Tr} (\rho_t \mathD [\{ \eta_k \}])$. 

In the interaction picture with respect to {the Hamiltonian (\ref{eq:Hamiltonian})}, henceforth denoted with a tilde, the equation of
motion for $\widetilde{\chi}_t [\{\eta_k \}]$ is given by
\begin{equation}
  \label{eq:equation-charfun} \partial_t \widetilde{\chi}_t [\{\eta_k \}] = - \Gamma_t [\{\eta_k \}]\, \widetilde{\chi}_t
[\{\eta_k \}].
\end{equation}
{ This follows from Eq.~(\ref{eq:master-equation}) using the cyclic property of the trace and the canonical commutation relations in Weyl form.}
For an isotropic $g(\xi)=g_r(|\xi|)/2\pi$ { the $\xi$ integral is conveniently calculated in
polar coordinates. Using the Jacobi-Anger formula for the Bessel function \cite{Abramowitz}} the decoherence rate
\begin{equation}
\label{eq:decoherenceRate}
\Gamma_t [\{\eta_k \}] = \gamma - \frac{\gamma}{L}
\int_0^L \mathd x \; \hat{g}_r(s_t[\{\eta_k\},x])
\end{equation}
can be written in terms of the Hankel transform of $g$,
$\hat{g}_r(s)=\int_0^\infty {\rm d}r \,r g_r(r) J_0(sr)$, evaluated at
\begin{equation}
  \label{eq:argR} s_t[\{\eta_k\},x] = 2 \left| \sum_{k \in K} \Omega^+_k f_k e^{\mathi kx}
e^{\mathi \omega_k t} \eta_k + \Omega^-_k f_k^{\ast} e^{- \mathi kx}
e^{- \mathi \omega_k t} \eta^{\ast}_k \right|.
\end{equation}

Equation {\eqref{eq:equation-charfun}} can be readily solved up to a
quadrature. It will be used below to analyze the dynamics of the
purity decay.
Moreover, it serves as the starting point to derive the semiclassical dynamics of the field. In order to do that, we first reformulate the above results in the language of functional calculus.

\section{Equation of motion for the Wigner functional}

The semiclassical field dynamics induced by Eq.~(\ref{eq:master-equation}) is best described in the phase space of the canonical field variables $\Phi$ and $\Pi$. In this representation, and using the complex field (\ref{eq:complexField}), the Weyl operators take the form $\mathD
[\eta] = \exp \left( \int_0^L \mathd x \, \left[\eta (x) \Psi^{\dagger} (x)
- \eta^{\ast} (x) \Psi (x)\right] \right)$. They are operator-valued functionals effecting a phase-space displacement of the canonical field variables at each point $x$ by the complex  wave amplitude $\eta(x)$. [Note that $\eta(x)$ has dimension of reciprocal square root of length, the same as $\Psi(x)$ and $\xi$.]

In this representation, the equation of motion for the characteristic functional takes a form analogous to  (\ref{eq:equation-charfun}), 
\begin{equation}
\label{eq:equation-charfun2} \partial_t \widetilde{\chi}_t [\eta] = - \Gamma_t [\eta]\, \widetilde{\chi}_t
[\eta].
\end{equation} 
Here, the decoherence rate is a functional of the complex wave amplitude $\eta(x)$,
\begin{equation}
\label{eq:decoherence-rate-position-basis}
\Gamma_t [\eta] = \gamma - \frac{\gamma}{L} \int_0^L \!\mathd x \int
\! \mathd^2 \xi \,g (\xi) e^{ -\int_0^L \mathd y [\eta^\ast (y) \Lambda_{t} (\xi,x;y) + \mathrm{c.c.}]}.
\end{equation}
It involves the interaction-picture phase-space displacements $\Lambda_{t}
(\xi,x;y) = \xi a_t (x ; y) - \xi^\ast b_t (x ; y)$, with 
\begin{subequations}
\begin{eqnarray}
a_t (x ; y) &=& \frac{1}{\sqrt{L}} \sum_{k\in K} 
\left [\cos \omega_k t - \frac{\mathi}{2} \left ( \frac{\omega}{\omega_k} + \frac{\omega_k}{\omega} \right ) \sin \omega_k t \right ] \nonumber \\
&&\times f_k e^{\mathi k (x-y)} \\
b_t (x ; y) &=& \frac{\mathi}{2\sqrt{L}} \sum_{k\in K}\!
\left ( \frac{\omega}{\omega_k} - \frac{\omega_k}{\omega} \right )\! \sin (\omega_k t)f_k e^{\mathi k (x-y)}.
\end{eqnarray}
\end{subequations}

The Wigner functional of the field state is defined as the functional Fourier transform of $\widetilde{\chi}_t[\eta]$	
\begin{equation} \label{eq:wigner}
	\widetilde{W}_t[\lambda] = \int {\cal D}^2 [\eta]\, \widetilde{\chi}_t[\eta] \exp \left (\int_0^L \! \mathd y \, [\lambda(y) \eta^\ast(y) - {\rm c.c}] \right),
\end{equation}
{ where the functional integral is defined as the limit $n \to \infty$ of the corresponding integral over $n$ mode variables \cite{Dalton}}.
The equation of motion for the Wigner functional follows from the Fourier transform of (\ref{eq:equation-charfun2}) as
\begin{eqnarray} \partial_t \widetilde{W}_t [\lambda] & =
& - \gamma \,\widetilde{W}_t[\lambda] + \frac{\gamma}{L} \! \int_0^L \!  \mathd x \! \int \! \mathd^2 \xi \,g
(\xi)\widetilde{W}_t [\lambda - \Lambda_t(\xi,x)].\nonumber\\
& & 
\label{eq:Boltzmann-equation}
\end{eqnarray}
This equation describes the time evolution of a quasiprobability distribution on a
functional phase space. Each point therein is described by a complex function $\lambda(y)$ corresponding to a linear combination of the canonical field variables. The latter are subject to random kicks whose strength is given by the function $\Lambda_t(\xi,x;y)$, playing the role of the random variable  $\xi f(y/L - x/L)$ in the interaction picture. 
It thus follows that 
Eq.~(\ref{eq:master-equation}) can  be considered  the field-theoretic  generalization of a quantum linear Boltzmann equation \cite{Vacchini}.

We note that in the diffusion limit of small and frequent kicks  Eq.~(\ref{eq:Boltzmann-equation}) can be approximated by a Fokker-Planck equation {\cite{Pawula}.
Expanding  the exponential in Eq.~(\ref{eq:decoherence-rate-position-basis}) 
to second order in $\xi$ and using that $g(\xi)$ is an even function, the dynamics of $\widetilde{\chi}_t[\eta]$ reduces to
\begin{equation}\label{eq:17}
\partial_t  \widetilde{\chi}_t [\eta] = - \frac{\gamma \sigma_g^2}{L} \!\int_0^L\!
   \mathd x \left| \int\! \mathd y \;[\eta (y) a_t (x ; y) + \eta^{\ast} (y) b_t
   (x ; y)] \right|^2,
\end{equation}
where $\sigma^2_g$ is the second moment of $g$.
The functional Fourier transform of (\ref{eq:17}) can be written in terms of functional derivatives of the Wigner functional using
\begin{eqnarray}
\frac{\delta}{\delta \lambda (y)}  \widetilde{W}_t [\lambda] &=& \int
   \mathcal{D}^2[\eta] \, \eta^{\ast} (y) \widetilde{\chi}_t [\eta] e^{ 
   \int \mathd z [\lambda (z) \eta^{\ast} (z) - \lambda^{\ast} (z) \eta (z)]},\nonumber\\
\frac{\delta}{\delta \lambda^{\ast} (y)}  \widetilde{W}_t [\lambda] &=& -
   \int \mathcal{D}^2[\eta] \, \eta (y) \widetilde{\chi}_t [\eta] e^{\int
   \mathd z [\lambda (z) \eta^{\ast} (z) - \lambda^{\ast} (z) \eta (z)]},\nonumber\\
\end{eqnarray}
which make use of the identity $\frac{\delta \lambda (z)}{\delta \lambda (y)} = \delta (z - y)$ \cite{Zeidler}. The resulting equation is}
\begin{eqnarray}
 \partial_t \widetilde{W}_t [\lambda] &=& \int \mathd x_1 \mathd x_2 \left[Q^{\lambda\lambda}_t(x_2 - x_1)\frac{\delta}{\delta \lambda (x_1) } \frac{\delta}{\delta\lambda (x_2)}\right.\nonumber\\
 &&+ Q^{\lambda\lambda^\ast}_t(x_2 - x_1)\frac{\delta}{\delta \lambda (x_1) } \frac{\delta}{\delta\lambda^\ast (x_2)}\nonumber\\
 && + Q^{\lambda^\ast\lambda}_t(x_2 - x_1)\frac{\delta}{\delta \lambda^\ast (x_1) } \frac{\delta}{\delta\lambda (x_2)}\nonumber\\
 && \left. + Q^{\lambda^\ast\lambda^\ast}_t(x_2 - x_1)\frac{\delta}{\delta \lambda^\ast (x_1) } \frac{\delta}{\delta\lambda^\ast (x_2)} \right]\widetilde{W}_t[\lambda],\nonumber\\
\end{eqnarray}
with the coefficients
\begin{eqnarray} 
Q^{\lambda\lambda}_t(x_2 - x_1) &=& -\gamma \sigma_g^2\int_0^L \! \frac{\mathd x}{L} b_t (x ; x_1) a^{\ast}_t (x ; x_2),   \notag \\
Q^{\lambda\lambda^\ast}_t(x_2 - x_1) &=& \gamma \sigma_g^2\int_0^L \! \frac{\mathd x}{L} b_t (x ; x_1) b^{\ast}_t (x ; x_2), \notag \\
Q^{\lambda^\ast\lambda}_t(x_2 - x_1) &=& \gamma \sigma_g^2\int_0^L \! \frac{\mathd x}{L} a_t (x ; x_1) a^{\ast}_t (x ; x_2), \notag \\
Q^{\lambda^\ast\lambda^\ast}_t(x_2 - x_1) &=&- \gamma \sigma_g^2\int_0^L \! \frac{\mathd x}{L} a_t (x ; x_1) b^{\ast}_t (x ; x_2).\nonumber\\
\end{eqnarray}

Since the Wigner functional \eqref{eq:wigner} is analytic in the complex wave amplitude $\lambda$, the functional derivatives treat $\lambda$ and $\lambda^\ast$ as independent variables, like in the Wirtinger calculus \cite{Fischer}. This field-theoretic Fokker-Planck equation can be solved using the techniques introduced in \cite{Graham}.

The Wigner representation is also convenient for describing the gradual loss of
quantumness induced by the exact Eq.~(\ref{eq:master-equation}). This is because quantum
superpositions of macroscopically distinct field states are characterized by a
quasiprobability distribution displaying strong oscillations between positive and
negative values in a local phase-space region of volume $\hbar$.
The decoherence of a superposition due to random phase-space kicks results in the
blurring of these fine structures. Once the Wigner functional is nonnegative, it can be
viewed as a probability distribution in a functional phase space. The corresponding
field state is then indistinguishable from a (mixed) classical field configuration.
This loss of coherence will now be quantified through the decay of the purity of the
state.

\section{Decoherence dynamics}
Physically, one would expect that the
purity decay rate of a quantum superposition of field states depends on
their initial separation as compared to the characteristic width of $g$. Moreover,
as we showed in Eq.~(\ref{eq:monotonic-purity-decay}), the purity of a superposition
decays monotonically with time. Its functional dependence on the parameters of
Eq.~(\ref{eq:master-equation}) is determined by the state of the field. 
In the following we characterize the dynamics of the purity decay both 
numerically and analytically for superpositions of single-mode coherent states in the
ground mode, $\ket{\psi} = \mathcal{N}(\ket{\alpha}_0 + \ket{\beta}_0) \ket{\{0\}_{k\neq 0}}$, {with $\mathcal{N} = \left[ 2 + 2 \, \exp (- \frac12 | \alpha - \beta |^2) \cos (2 \tmop{Im} \alpha
\beta^{\ast}) \right]^{- 1}$}.
\begin{figure}[t]
	\includegraphics[width=245pt]{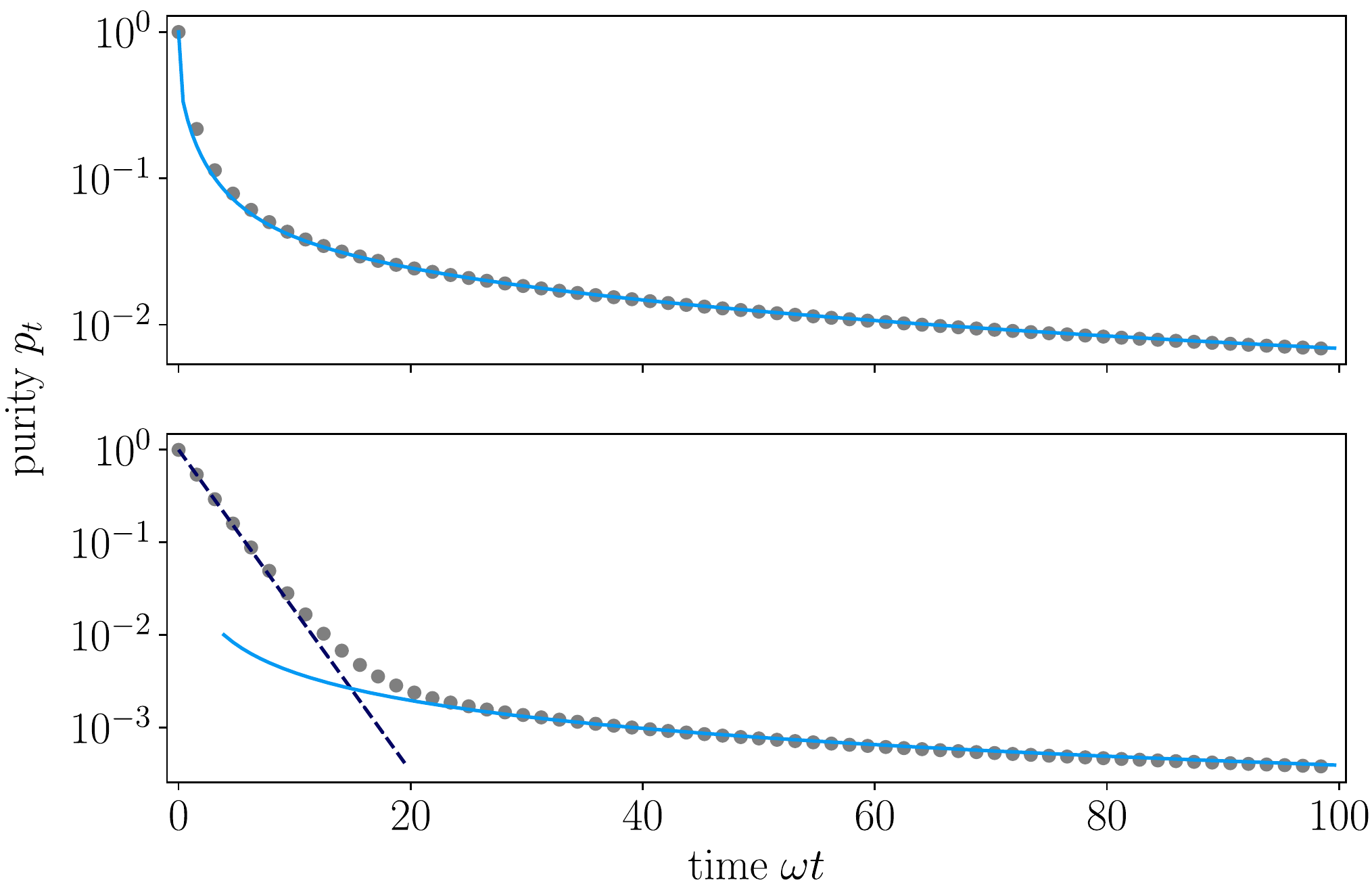}
	\caption{\label{fig:purity-decay-with-time} (Color online) Purity decay of a
		superposition of single-mode coherent states  $\ket{\psi} = \mathcal{N} (\ket{\alpha}_0 + \ket{-\alpha}_0)\ket{\{0\}_{k\neq 0}}$, with $\alpha = 2 + 2\mathi$, $v/L\omega=0.01$, and $\sigma_x/L = 1$. The dots correspond to the exact numerical calculation of Eq.~(\ref{eq:purity}).
		The top panel shows the case of a narrow kick distribution with $\sigma^2_g L=0.32$ and for $\gamma/\omega = 1$. The solid line is the analytic approximation of Eq.~\eqref{eq:purity} for $\sigma^2_g L \ll 1$ and $\sigma_x/L \simeq 1$ { given in Eq.~(\ref{eq:purity-broad-f-narrow-g})}.
		The opposite case of a broad kick distribution with $\sigma^2_g L = 32$ and for $\gamma/\omega = 0.2$ is illustrated in the bottom panel. 
		The solid line gives the long-time behavior obtained using the Laplace approximation, $p_t \simeq 1/4 \gamma t L \sigma_g^2$, while the dashed line shows the short-time behavior $p_t\simeq \exp(-2\gamma t)$.
}
\end{figure}

The purity of the time-evolved quantum field in the mode representation follows from {the solution to}
Eq.~\eqref{eq:equation-charfun},
\begin{eqnarray}
\label{eq:purity} p_t &=& \int {\cal D}^2[\{ \eta_k \}]\, |
\chi_0 [\{\eta_{k} \}] |^2 \exp \left( - 2
\int_0^t \mathd \tau \Gamma_{\tau} [\{\eta_{k} \}] \right),
\nonumber\\
\end{eqnarray}
where $\chi_0$ is the characteristic functional of the
initial state.
For definiteness, we take the kick distribution to be given by
\begin{equation}
\label{eq:kick-distribution}
g(\xi)=\exp(- \vert \xi \vert^2 / 2 \sigma_g^2)
/2\pi \sigma_g^2,
\end{equation}
and choose the spread function $f(s)$ so that
\begin{equation}
\label{eq:fk}
f_k=\sqrt{L}\exp(-\sigma_x^2k^2/2)/\vartheta_3{\boldsymbol(}0,\exp(-2\pi^2 \sigma_x^2/L ^2)\boldsymbol{)},
\end{equation}
where $\vartheta_3$ is the Jacobi theta function \cite{Abramowitz}. In the following we consider a broad spread function with $\sigma_x/L \simeq 1$. In this case the approximations $f_k \simeq \sqrt{L} \delta_{k,
0}$ and $\Gamma_t (\{\eta_k \}) \simeq \gamma - \gamma \exp (- 2 L
\sigma_g^2 | \eta_0 |^2)$ can be used.

For the numerical calculation of Eq.~\eqref{eq:purity} the quantum field is modeled as
a harmonic chain of 32 local oscillators. The corresponding phase space is discretized
using a generalized Faure sequence {\cite{LEcuyer,Papageorgiou,Owen}}.

Figure \ref{fig:purity-decay-with-time} shows the purity decay for two
limiting values of the width of the kick distribution. The initial state is given by
$\ket{\psi} = \mathcal{N} (\ket{\alpha}_0 + \ket{-\alpha}_0) \ket{\{0\}_{k\neq 0}}$. 
In the limit of a narrow distribution $g$, such that $\sigma^2_g L \ll 1$, the purity can be calculated analytically from Eq.~\eqref{eq:purity},
{\begin{eqnarray}
\label{eq:purity-broad-f-narrow-g}
p_t &\simeq& \frac{2\mathcal{N}^2}{1 + \mu^2} \left[ 1 + e^{- \frac{| \alpha - \beta |^2}{1 + \mu^2}} + 4 e^{- \frac{1}{2} | \alpha - \beta |^2} \cos \left( \mathrm{Im} \alpha
\beta^{\ast} \right) \right.\nonumber\\
&&\left. + e^{- \frac{| \alpha -
\beta |^2 \mu^2}{1 + \mu^2}} + e^{- | \alpha - \beta |^2} \cos \left( 2 \mathrm{Im}
\alpha \beta^{\ast} \right) \right],
\end{eqnarray}
where $\mu^2 = 4\gamma t \sigma_g^2 L$}. This expression corresponds to the solid line in the top panel of Fig.~\ref{fig:purity-decay-with-time}, which is in excellent agreement with the exact numerical calculation of Eq.~\eqref{eq:purity}. For the case of a broad distribution, $\sigma^2_g L \gg 1$, the purity cannot be calculated analytically for all times. Its short-time behavior is given by $p_t \simeq \exp(-2\gamma t)$, as indicated by the dashed line in the bottom panel, and its long-time evolution can be determined using Laplace's method {\cite{Bleistein}}, yielding $p_t \sim 1/4 \gamma t L \sigma_g^2$ as $\gamma t\to \infty$. This asymptotic behavior is indicated by the solid curve in the bottom panel of Fig.~\ref{fig:purity-decay-with-time}.

\begin{figure}[t]
	\includegraphics[width=245pt]{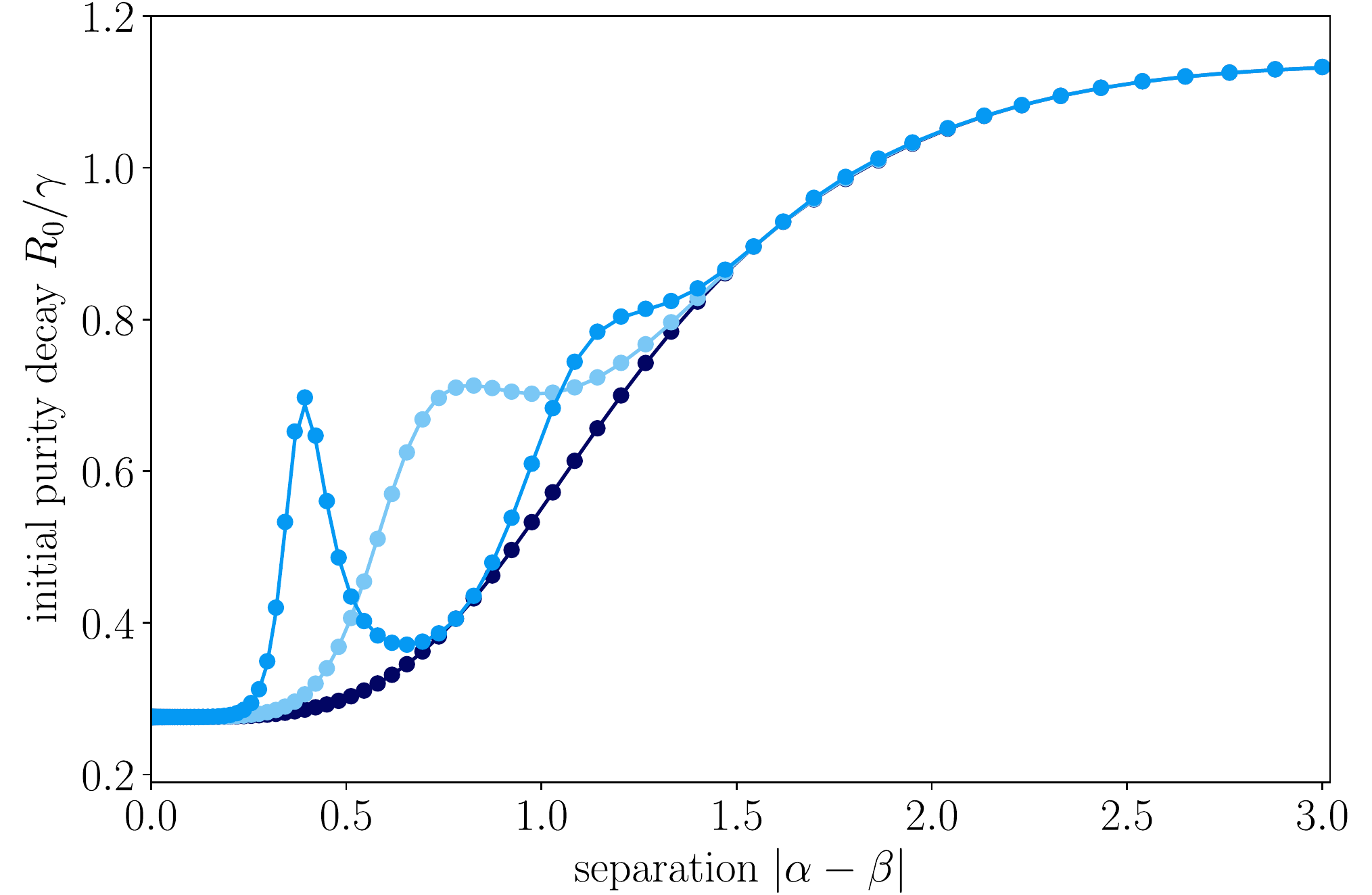}
	\caption{\label{fig:purity-decay-moving-center} (Color online) Initial purity decay
		rate, $R_0/\gamma$, of a superposition of single-mode coherent states $\ket{\psi} = \mathcal{N} (\ket{\alpha}_0 + \ket{\beta}_0) \ket{\{0\}_{k\neq 0}}$ as a function
		of the separation $|\alpha - \beta|$, for the case $\sigma_x/L = 1$, with $v/L\omega = 0.01$ and $\gamma/\omega = 1.0$. The curves show the behavior of $R_0/\gamma$ for a narrow kick distribution with $\sigma_g^2 L = 0.08$. They correspond to superpositions such that $\alpha+\beta=8,4,0$ (from left to right) with $\alpha - \beta=i|\alpha-\beta|$. The circles give the  exact numerical values of Eq.~\eqref{eq:initial-purity-decay}, and the solid lines show the analytic approximation of this equation for $\sigma_x/L \simeq 1$.}
\end{figure}

In order to investigate how the purity decay depends on the separation $\vert \alpha - \beta \vert$ between the superposed coherent states, we calculate the initial purity decay rate $R_0=-\dot{p}_0$ from Eq.~\eqref{eq:purity},
\begin{equation}
\label{eq:initial-purity-decay}
R_0 = 2 \int{\cal D}^2[\{\eta_k\}] \, | \chi_0 [\{\eta_k \}] |^2 \, \Gamma_0 [\{\eta_{k} \}],
\end{equation}
for different initial states. For a broad spread function with $\sigma_x/L \simeq 1$ the phase-space
integral can be expressed analytically. One finds that $R_0 - 2\gamma$
equals Eq.~(\ref{eq:purity-broad-f-narrow-g}) multiplied by $-2\gamma$, with $\mu^2 = 2\sigma_g^2L$. 

Figure \ref{fig:purity-decay-moving-center} shows that in general $R_0$ does not
increase monotonically with the separation $|\alpha - \beta|$; it exhibits oscillations for
$|\alpha - \beta| < 1.5$. Moreover, for large separations the
decoherence rate approaches the maximum value 
\begin{equation}
\frac{R_{\rm max}}{\gamma} = 2 -\frac{1}{1 + 2\sigma^2_g L}.
\end{equation}
In the case of a broad kick distribution $R_0$ does not vary
appreciably with the separation and approaches the value
$R_{\rm max}/\gamma=2$ (not shown)}. We note that the analytic expression
for Eq.~(\ref{eq:initial-purity-decay}) (solid curves), calculated assuming $\sigma_x/L \simeq 1$,
 is in excellent agreement with the numerical
calculation (circles) of the exact decay rate $R_0$.

\section{Mean energy increase}
 In addition to inducing decoherence, a quantum master equation will in general also affect the dynamics of 
otherwise conserved quantities such as energy \cite{Bassi}. Experimental bounds on 
the observed conservation of energy will therefore constrain the parameters entering
the non-unitary time evolution.

{From the master equation (\ref{eq:master-equation}) it follows that the field energy increases with a constant rate that is independent of the quantum state of the field,
\begin{equation}
\label{energy-expectation} \partial_t \left \langle \Eta \right \rangle_t = \gamma \sum_{k\in K}\hbar \omega_k \vert f_k \vert^2 \int \mathd^2 \xi \, g(\xi)  \vert \xi \Omega^+_k - \xi^{\ast} \Omega^-_k \vert^2.
\end{equation}
This is obtained by using the exact expansion
$e^{-\mathrm{X}} \mathrm{H} e^{\mathrm{X}} = \mathrm{H} - \left[
\mathrm{X}, \mathrm{H} \right] + \frac{1}{2} \left[ \mathrm{X}, \left[
\mathrm{X}, \mathrm{H} \right] \right]$, where $ \mathrm{X}$ is the exponent in Eq.~(\ref{eq:displacement-op}).

In the limit of large $L$ { the sum over $k$ can be approximated as an integral and the heating rate has the explicit expression}
\begin{equation}
\label{energy-expectation2} \partial_t \left\langle \Eta \right\rangle_t =
 2\sqrt{\pi} \gamma \hbar \omega  \sigma_x \sigma_g^2 \left( 1 + \frac{ v^2}{(2 \sigma_x \omega)^2} \right),
\end{equation}
where we used the distribution (\ref{eq:kick-distribution}) and the spread function (\ref{eq:fk}).

\section{Conclusions}

We introduced a generic Lindblad master
equation that serves to decohere a nonrelativistic bosonic field. The Ehrenfest equations for the canonical field variables remain identical
with the corresponding classical field equations, while quantum superpositions
of distinct field configurations are rapidly turned into mixtures. In fact,
the master equation induces a monotonic decay of the purity.

We showed that the Wigner functional is an appropriate representation
to capture the gradual quantum-to-classical transition of the field.
To the extent to which the functional turns positive, its dynamics can
be regarded as being governed by a linear Boltzmann equation, which in
the diffusion limit reduces to a Fokker-Planck
equation. Using the characteristic functional, the
decoherence rate of a quantum superposition of two effectively
classical field configurations was shown to depend nontrivially on
their phase-space separation, and to saturate for large separations.

The effect of the master equation on the field may be viewed as
arising from a stochastic process of generalized simultaneous
measurements of the field amplitude and its canonical momentum, whose
outcomes are discarded. These fictitious measurements have finite spatial
resolution, as characterized by the spread function $f$. It is important
to remark that only due to the finite width $\sigma_x>0$ of $f$
in position space is the continuous field dynamics  physically consistent and
divergence-free. Moreover, the limited resolution $\sigma_g>0$
associated with the generalized measurement of the phase-space
coordinates  ensures a finite back-action on all canonical field
variables.

Notwithstanding the generality of the master equation and its complex
decoherence dynamics, several analytical expressions were
obtained for functionals of the field state. In particular, we obtained 
the functional dependence of the purity in the most important limiting cases,
and we calculated the exact energy increase. The analytical
results are in remarkable agreement with numerical calculations. This
shows that for a superposition of field coherent states expectation values 
can be accurately calculated using quasi Monte Carlo integration based
on generalized Faure sequences, despite the high-dimensional character of 
the phase space.

We note that in Ref.~{\cite{Shimizu}} the stability of the quantum
state of a macroscopic number of degrees of freedom against
perturbation by a quantum or a classical noise was analyzed based on
general considerations. Since the present work introduces a concrete
decoherence model, it should encourage further investigations of
macroscopic quantum systems described by quantum fields.

The methods discussed in this work can be straightforwardly
generalized to non-relativistic bosonic tensor fields. In principle, a
similar treatment can be developed for fermionic fields, even though
their classical analog is less evident. Finally, a relativistic
generalization of the model presented here would enable the study of
the quantum-to-classical transition of quantum electrodynamics.

M.B. acknowledges financial support from CONACYT (Mexico) and DAAD (Germany).

\end{document}